

%
%
%
\def\unredoffs{} \def\redoffs{\voffset=-.31truein\hoffset=-.59truein}
\def\speclscape{\special{ps: landscape}}
%
%
%
%
\newbox\leftpage \newdimen\fullhsize \newdimen\hstitle \newdimen\hsbody
\tolerance=1000\hfuzz=2pt
\catcode`\@=11 
\def\bigans{b }
\def\answ{b }

%
\ifx\answ\bigans\message{(This will come out unreduced.}
\magnification=1200\unredoffs\baselineskip=16pt plus 2pt minus 1pt
\hsbody=\hsize \hstitle=\hsize 
\else\message{(This will be reduced.} \let\l@r=L
\magnification=1000\baselineskip=16pt plus 2pt minus 1pt \vsize=7truein
\redoffs \hstitle=8truein\hsbody=4.75truein\fullhsize=10truein\hsize=\hsbody
\output={\ifnum\pageno=0 
  \shipout\vbox{\speclscape{\hsize\fullhsize\makeheadline}
    \hbox to \fullhsize{\hfill\pagebody\hfill}}\advancepageno
  \else
  \almostshipout{\leftline{\vbox{\pagebody\makefootline}}}\advancepageno
  \fi}
\def\almostshipout#1{\if L\l@r \count1=1 \message{[\the\count0.\the\count1]}
      \global\setbox\leftpage=#1 \global\let\l@r=R
 \else \count1=2
  \shipout\vbox{\speclscape{\hsize\fullhsize\makeheadline}
      \hbox to\fullhsize{\box\leftpage\hfil#1}}  \global\let\l@r=L\fi}
\fi
%
\newcount\yearltd\yearltd=\year\advance\yearltd by -1900

\def\Title#1#2{\nopagenumbers\abstractfont\hsize=\hstitle\rightline{#1}%
\vskip 1in\centerline{\titlefont #2}\abstractfont\vskip .5in\pageno=0}
\def\Date#1{\vfill\leftline{#1}\tenpoint\supereject\global\hsize=\hsbody%
\footline={\hss\tenrm\folio\hss}}
%

\def\draftmode{\message{ DRAFTMODE }\def\draftdate{{\rm preliminary draft:
\number\month/\number\day/\number\yearltd\ \ \hourmin}}%
\headline={\hfil\draftdate}\writelabels\baselineskip=20pt plus 2pt minus 2pt
 {\count255=\time\divide\count255 by 60 \xdef\hourmin{\number\count255}
  \multiply\count255 by-60\advance\count255 by\time
  \xdef\hourmin{\hourmin:\ifnum\count255<10 0\fi\the\count255}}}
\def\nolabels{\def\wrlabeL##1{}\def\eqlabeL##1{}\def\reflabeL##1{}}
\def\writelabels{\def\wrlabeL##1{\leavevmode\vadjust{\rlap{\smash%
{\line{{\escapechar=` \hfill\rlap{\sevenrm\hskip.03in\string##1}}}}}}}%
\def\eqlabeL##1{{\escapechar-1\rlap{\sevenrm\hskip.05in\string##1}}}%
\def\reflabeL##1{\noexpand\llap{\noexpand\sevenrm\string\string\string##1}}}
\nolabels
%
\global\newcount\secno \global\secno=0
\global\newcount\meqno \global\meqno=1
\def\newsec#1{\global\advance\secno by1\message{(\the\secno. #1)}
\global\subsecno=0\eqnres@t\noindent{\bf\the\secno. #1}
\writetoca{{\secsym} {#1}}\par\nobreak\medskip\nobreak}
\def\eqnres@t{\xdef\secsym{\the\secno.}\global\meqno=1\bigbreak\bigskip}
\def\sequentialequations{\def\eqnres@t{\bigbreak}}\xdef\secsym{}
\global\newcount\subsecno \global\subsecno=0
\def\subsec#1{\global\advance\subsecno by1\message{(\secsym\the\subsecno. #1)}
\ifnum\lastpenalty>9000\else\bigbreak\fi
\noindent{\it\secsym\the\subsecno. #1}\writetoca{\string\quad
{\secsym\the\subsecno.} {#1}}\par\nobreak\medskip\nobreak}
\def\appendix#1#2{\global\meqno=1\global\subsecno=0\xdef\secsym{\hbox{#1.}}
\bigbreak\bigskip\noindent{\bf Appendix #1. #2}\message{(#1. #2)}
\writetoca{Appendix {#1.} {#2}}\par\nobreak\medskip\nobreak}
%
%
\def\eqnn#1{\xdef #1{(\secsym\the\meqno)}\writedef{#1\leftbracket#1}%
\global\advance\meqno by1\wrlabeL#1}
\def\eqna#1{\xdef #1##1{\hbox{$(\secsym\the\meqno##1)$}}
\writedef{#1\numbersign1\leftbracket#1{\numbersign1}}%
\global\advance\meqno by1\wrlabeL{#1$\{\}$}}
\def\eqn#1#2{\xdef #1{(\secsym\the\meqno)}\writedef{#1\leftbracket#1}%
\global\advance\meqno by1$$#2\eqno#1\eqlabeL#1$$}
%
\newskip\footskip\footskip14pt plus 1pt minus 1pt 
\def\footnotefont{\ninepoint}\def\f@t#1{\footnotefont #1\@foot}
\def\f@@t{\baselineskip\footskip\bgroup\footnotefont\aftergroup\@foot\let\next}
\setbox\strutbox=\hbox{\vrule height9.5pt depth4.5pt width0pt}
\global\newcount\ftno \global\ftno=0
\def\foot{\global\advance\ftno by1\footnote{$^{\the\ftno}$}}
%
\newwrite\ftfile
\def\footend{\def\foot{\global\advance\ftno by1\chardef\wfile=\ftfile
$^{\the\ftno}$\ifnum\ftno=1\immediate\openout\ftfile=foots.tmp\fi%
\immediate\write\ftfile{\noexpand\smallskip%
\noexpand\item{f\the\ftno:\ }\pctsign}\findarg}%
\def\footatend{\vfill\eject\immediate\closeout\ftfile{\parindent=20pt
\centerline{\bf Footnotes}\nobreak\bigskip\input foots.tmp }}}
\def\footatend{}
%
%
\global\newcount\refno \global\refno=1
\newwrite\rfile
\def\ref{[\the\refno]\nref}
\def\nref#1{\xdef#1{[\the\refno]}\writedef{#1\leftbracket#1}%
\ifnum\refno=1\immediate\openout\rfile=refs.tmp\fi
\global\advance\refno by1\chardef\wfile=\rfile\immediate
\write\rfile{\noexpand\item{#1\ }\reflabeL{#1\hskip.31in}\pctsign}\findarg}
\def\findarg#1#{\begingroup\obeylines\newlinechar=`\^^M\pass@rg}
{\obeylines\gdef\pass@rg#1{\writ@line\relax #1^^M\hbox{}^^M}%
\gdef\writ@line#1^^M{\expandafter\toks0\expandafter{\striprel@x #1}%
\edef\next{\the\toks0}\ifx\next\em@rk\let\next=\endgroup\else\ifx\next\empty%
\else\immediate\write\wfile{\the\toks0}\fi\let\next=\writ@line\fi\next\relax}}
\def\striprel@x#1{} \def\em@rk{\hbox{}}
\def\lref{\begingroup\obeylines\lr@f}
\def\lr@f#1#2{\gdef#1{\ref#1{#2}}\endgroup\unskip}

\def\addref#1{\immediate\write\rfile{\noexpand\item{}#1}} 
\def\startrefs#1{\immediate\openout\rfile=refs.tmp\refno=#1}
\def\xref{\expandafter\xr@f}\def\xr@f[#1]{#1}
\def\refs#1{\count255=1[\r@fs #1{\hbox{}}]}
\def\r@fs#1{\ifx\und@fined#1\message{reflabel \string#1 is undefined.}%
\nref#1{need to supply reference \string#1.}\fi%
\vphantom{\hphantom{#1}}\edef\next{#1}\ifx\next\em@rk\def\next{}%
\else\ifx\next#1\ifodd\count255\relax\xref#1\count255=0\fi%
\else#1\count255=1\fi\let\next=\r@fs\fi\next}
%

%
\newwrite\ffile\global\newcount\figno \global\figno=1
\def\fig{fig.~\the\figno\nfig}
\def\nfig#1{\xdef#1{fig.~\the\figno}%
\writedef{#1\leftbracket fig.\noexpand~\the\figno}%
\ifnum\figno=1\immediate\openout\ffile=figs.tmp\fi\chardef\wfile=\ffile%
\immediate\write\ffile{\noexpand\medskip\noexpand\item{Fig.\ \the\figno. }
\reflabeL{#1\hskip.55in}\pctsign}\global\advance\figno by1\findarg}
\def\xfig{\expandafter\xf@g}\def\xf@g fig.\penalty\@M\ {}
\def\figs#1{figs.~\f@gs #1{\hbox{}}}
\def\f@gs#1{\edef\next{#1}\ifx\next\em@rk\def\next{}\else
\ifx\next#1\xfig #1\else#1\fi\let\next=\f@gs\fi\next}
\newwrite\lfile
{\escapechar-1\xdef\pctsign{\string\%}\xdef\leftbracket{\string\{}
\xdef\rightbracket{\string\}}\xdef\numbersign{\string\#}}

\def\writestop{\def\writestoppt{\immediate\write\lfile{\string\pageno%
\the\pageno\string\startrefs\leftbracket\the\refno\rightbracket%
\string\def\string\secsym\leftbracket\secsym\rightbracket%
\string\secno\the\secno\string\meqno\the\meqno}\immediate\closeout\lfile}}
\def\writestoppt{}\def\writedef#1{}
\def\seclab#1{\xdef #1{\the\secno}\writedef{#1\leftbracket#1}\wrlabeL{#1=#1}}
\def\subseclab#1{\xdef #1{\secsym\the\subsecno}%
\writedef{#1\leftbracket#1}\wrlabeL{#1=#1}}
\newwrite\tfile \def\writetoca#1{}
\def\leaderfill{\leaders\hbox to 1em{\hss.\hss}\hfill}
\def\writetoc{\immediate\openout\tfile=toc.tmp
   \def\writetoca##1{{\edef\next{\write\tfile{\noindent ##1
   \string\leaderfill {\noexpand\number\pageno} \par}}\next}}}

%
\catcode`\@=12 
%
\edef\tfontsize{\ifx\answ\bigans scaled\magstep3\else scaled\magstep4\fi}
\font\titlerm=cmr10 \tfontsize \font\titlerms=cmr7 \tfontsize
\font\titlermss=cmr5 \tfontsize \font\titlei=cmmi10 \tfontsize
\font\titleis=cmmi7 \tfontsize \font\titleiss=cmmi5 \tfontsize
\font\titlesy=cmsy10 \tfontsize \font\titlesys=cmsy7 \tfontsize
\font\titlesyss=cmsy5 \tfontsize \font\titleit=cmti10 \tfontsize
\skewchar\titlei='177 \skewchar\titleis='177 \skewchar\titleiss='177
\skewchar\titlesy='60 \skewchar\titlesys='60 \skewchar\titlesyss='60
\def\titlefont{\def\rm{\fam0\titlerm}
\textfont0=\titlerm \scriptfont0=\titlerms \scriptscriptfont0=\titlermss
\textfont1=\titlei \scriptfont1=\titleis \scriptscriptfont1=\titleiss
\textfont2=\titlesy \scriptfont2=\titlesys \scriptscriptfont2=\titlesyss
\textfont\itfam=\titleit \def\it{\fam\itfam\titleit}\rm}
 \ifx\answ\bigans\else scaled\magstep1\fi
\ifx\answ\bigans\def\abstractfont{\tenpoint}\else
\font\abssl=cmsl10 scaled \magstep1
\font\absrm=cmr10 scaled\magstep1 \font\absrms=cmr7 scaled\magstep1
\font\absrmss=cmr5 scaled\magstep1 \font\absi=cmmi10 scaled\magstep1
\font\absis=cmmi7 scaled\magstep1 \font\absiss=cmmi5 scaled\magstep1
\font\abssy=cmsy10 scaled\magstep1 \font\abssys=cmsy7 scaled\magstep1
\font\abssyss=cmsy5 scaled\magstep1 \font\absbf=cmbx10 scaled\magstep1
\skewchar\absi='177 \skewchar\absis='177 \skewchar\absiss='177
\skewchar\abssy='60 \skewchar\abssys='60 \skewchar\abssyss='60
\def\abstractfont{\def\rm{\fam0\absrm}
\textfont0=\absrm \scriptfont0=\absrms \scriptscriptfont0=\absrmss
\textfont1=\absi \scriptfont1=\absis \scriptscriptfont1=\absiss
\textfont2=\abssy \scriptfont2=\abssys \scriptscriptfont2=\abssyss
\textfont\itfam=\bigit \def\it{\fam\itfam\bigit}\def\footnotefont{\tenpoint}%
\textfont\slfam=\abssl \def\sl{\fam\slfam\abssl}%
\textfont\bffam=\absbf \def\bf{\fam\bffam\absbf}\rm}\fi
\def\tenpoint{\def\rm{\fam0\tenrm}
\textfont0=\tenrm \scriptfont0=\sevenrm \scriptscriptfont0=\fiverm
\textfont1=\teni  \scriptfont1=\seveni  \scriptscriptfont1=\fivei
\textfont2=\tensy \scriptfont2=\sevensy \scriptscriptfont2=\fivesy
\textfont\itfam=\tenit \def\it{\fam\itfam\tenit}\def\footnotefont{\ninepoint}%
\textfont\bffam=\tenbf \def\bf{\fam\bffam\tenbf}\def\sl{\fam\slfam\tensl}\rm}
\font\ninerm=cmr9 \font\sixrm=cmr6 \font\ninei=cmmi9 \font\sixi=cmmi6
\font\ninesy=cmsy9 \font\sixsy=cmsy6 \font\ninebf=cmbx9
\font\nineit=cmti9 \font\ninesl=cmsl9 \skewchar\ninei='177
\skewchar\sixi='177 \skewchar\ninesy='60 \skewchar\sixsy='60
\def\ninepoint{\def\rm{\fam0\ninerm}
\textfont0=\ninerm \scriptfont0=\sixrm \scriptscriptfont0=\fiverm
\textfont1=\ninei \scriptfont1=\sixi \scriptscriptfont1=\fivei
\textfont2=\ninesy \scriptfont2=\sixsy \scriptscriptfont2=\fivesy
\textfont\itfam=\ninei \def\it{\fam\itfam\nineit}\def\sl{\fam\slfam\ninesl}%
\textfont\bffam=\ninebf \def\bf{\fam\bffam\ninebf}\rm}
%
%

\hyphenation{anom-aly anom-alies coun-ter-term coun-ter-terms}
\def\inv{^{\raise.15ex\hbox{${\scriptscriptstyle -}$}\kern-.05em 1}}

\def\Dsl{\,\raise.15ex\hbox{/}\mkern-13.5mu D} 
\def\dsl{\raise.15ex\hbox{/}\kern-.57em\partial}

\font\bigit=cmti10 scaled \magstep1
\def\lspace{\ifx\answ\bigans{}\else\qquad\fi}
\def\lbspace{\ifx\answ\bigans{}\else\hskip-.2in\fi} 
\def\boxeqn#1{\vcenter{\vbox{\hrule\hbox{\vrule\kern3pt\vbox{\kern3pt
	\hbox{${\displaystyle #1}$}\kern3pt}\kern3pt\vrule}\hrule}}}
\def\mbox#1#2{\vcenter{\hrule \hbox{\vrule height#2in
		\kern#1in \vrule} \hrule}}  
%
 \def\CC{{\cal C}}

\def\darr#1{\raise1.5ex\hbox{$\leftrightarrow$}\mkern-16.5mu #1}

\def\roughly#1{\raise.3ex\hbox{$#1$\kern-.75em\lower1ex\hbox{$\sim$}}}

\let\includefigures=\iftrue
\let\useblackboard=\iftrue
\newfam\black

\includefigures
\message{If you do not have epsf.tex (to include figures),}
\message{change the option at the top of the tex file.}
\input epsf
\def\figin{\epsfcheck\figin}\def\figins{\epsfcheck\figins}
\def\epsfcheck{\ifx\epsfbox\UnDeFiNeD
\message{(NO epsf.tex, FIGURES WILL BE IGNORED)}
\gdef\figin##1{\vskip2in}\gdef\figins##1{\hskip.5in}
\else\message{(FIGURES WILL BE INCLUDED)}%
\gdef\figin##1{##1}\gdef\figins##1{##1}\fi}
\def\DefWarn#1{}
\def\figinsert{\goodbreak\midinsert}
\def\ifig#1#2#3{\DefWarn#1\xdef#1{fig.~\the\figno}
\writedef{#1\leftbracket fig.\noexpand~\the\figno}%
\figinsert\figin{\centerline{#3}}\medskip\centerline{\vbox{
\baselineskip12pt\advance\hsize by -1truein
\noindent\footnotefont{\bf Fig.~\the\figno:} #2}}
\endinsert\global\advance\figno by1}
\else
\def\ifig#1#2#3{\xdef#1{fig.~\the\figno}
\writedef{#1\leftbracket fig.\noexpand~\the\figno}%
\global\advance\figno by1} \fi

\def\id{{1 \kern-.28em {\rm l}}}

\def\K3{{\bf K3}}
\def\journal#1&#2(#3){\unskip, \sl #1\ \bf #2 \rm(19#3) }
\def\andjournal#1&#2(#3){\sl #1~\bf #2 \rm (19#3) }

\def\hat{\widehat}

\def\frac#1#2{{#1\over#2}}

\def\inbar{\,\vrule height1.5ex width.4pt depth0pt}
\def\IC{\relax\hbox{$\inbar\kern-.3em{\rm C}$}}
\def\IR{\relax{\rm I\kern-.18em R}}
\def\IP{\relax{\rm I\kern-.18em P}}

%
%

%
\catcode`\@=11
\def\slash#1{\mathord{\mathpalette\c@ncel{#1}}}
\overfullrule=0pt

\def\CC{{\cal C}}

\def\II{{\cal I}}

\def\NN{{\cal N}}

\def\II{{\cal I}}

\def\underrel#1\over#2{\mathrel{\mathop{\kern\z@#1}\limits_{#2}}}

\catcode`\@=12


%


\def\p{{\partial}}

\def\ra{{\rightarrow}}



\lref\BriganteNU{
  M.~Brigante, H.~Liu, R.~C.~Myers, S.~Shenker and S.~Yaida,
  ``Viscosity Bound Violation in Higher Derivative Gravity,''
  Phys.\ Rev.\  D {\bf 77}, 126006 (2008)
  [arXiv:0712.0805 [hep-th]];
  ``The Viscosity Bound and Causality Violation,''
  Phys.\ Rev.\ Lett.\  {\bf 100}, 191601 (2008)
  [arXiv:0802.3318 [hep-th]].
}

\lref\HofmanMaldacena{
  D.~M.~Hofman and J.~Maldacena,
  ``Conformal collider physics: Energy and charge correlations,''
  JHEP {\bf 0805}, 012 (2008)
  [arXiv:0803.1467 [hep-th]].
}

\lref\BuchelTT{
  A.~Buchel and R.~C.~Myers,
  ``Causality of Holographic Hydrodynamics,''
  JHEP {\bf 0908}, 016 (2009)
  [arXiv:0906.2922 [hep-th]].
}

\lref\HofmanUG{
  D.~M.~Hofman,
  ``Higher Derivative Gravity, Causality and Positivity of Energy in a UV
  complete QFT,''
  Nucl.\ Phys.\  B {\bf 823}, 174 (2009)
  [arXiv:0907.1625 [hep-th]].
}

\lref\deBoerPN{
  J.~de Boer, M.~Kulaxizi and A.~Parnachev,
  ``$AdS_7/CFT_6$, Gauss-Bonnet Gravity, and Viscosity Bound,''
  JHEP {\bf 1003}, 087 (2010)
  [arXiv:0910.5347 [hep-th]].
}

\lref\CamanhoVW{
  X.~O.~Camanho and J.~D.~Edelstein,
  ``Causality constraints in AdS/CFT from conformal collider physics and
  Gauss-Bonnet gravity,''
  JHEP {\bf 1004}, 007 (2010)
  [arXiv:0911.3160 [hep-th]].
}

\lref\BuchelSK{
  A.~Buchel, J.~Escobedo, R.~C.~Myers, M.~F.~Paulos, A.~Sinha and M.~Smolkin,
  ``Holographic GB gravity in arbitrary dimensions,''
  JHEP {\bf 1003}, 111 (2010)
  [arXiv:0911.4257 [hep-th]].
}

\lref\deBoerGX{
  J.~de Boer, M.~Kulaxizi and A.~Parnachev,
  ``Holographic Lovelock Gravities and Black Holes,''
  arXiv:0912.1877 [hep-th].
}

\lref\CamanhoHU{
  X.~O.~Camanho and J.~D.~Edelstein,
  ``Causality in AdS/CFT and Lovelock theory,''
  arXiv:0912.1944 [hep-th].
}

\lref\MyersJV{
  R.~C.~Myers, M.~F.~Paulos and A.~Sinha,
  ``Holographic studies of quasi-topological gravity,''
  arXiv:1004.2055 [hep-th].
}

\lref\MyersRU{
  R.~C.~Myers and B.~Robinson,
  ``Black Holes in Quasi-topological Gravity,''
  arXiv:1003.5357 [gr-qc].
}

\lref\OsbornCR{
  H.~Osborn and A.~C.~Petkou,
  ``Implications of Conformal Invariance in Field Theories for General
  Dimensions,''
  Annals Phys.\  {\bf 231}, 311 (1994)
  [arXiv:hep-th/9307010].
}

\lref\KovtunEV{
  P.~K.~Kovtun and A.~O.~Starinets,
  ``Quasinormal modes and holography,''
  Phys.\ Rev.\  D {\bf 72}, 086009 (2005)
  [arXiv:hep-th/0506184].
}

\lref\LatorreEA{
  J.~I.~Latorre and H.~Osborn,
  ``Modified weak energy condition for the energy momentum tensor in  quantum
  field theory,''
  Nucl.\ Phys.\  B {\bf 511}, 737 (1998)
  [arXiv:hep-th/9703196].
}

\lref\ParnachevYT{
  A.~Parnachev and S.~S.~Razamat,
  ``Comments on Bounds on Central Charges in N=1 Superconformal Theories,''
  JHEP {\bf 0907}, 010 (2009)
  [arXiv:0812.0781 [hep-th]].
}

\lref\OlivaEB{
  J.~Oliva and S.~Ray,
  ``A new cubic theory of gravity in five dimensions: Black hole, Birkhoff's
  theorem and C-function,''
  arXiv:1003.4773 [gr-qc].
}

\lref\Anselmia{
  D.~Anselmi,
  ``The N = 4 quantum conformal algebra,''
  Nucl.\ Phys.\  B {\bf 541}, 369 (1999)
  [arXiv:hep-th/9809192].
}

\lref\KulaxiziPZ{
  M.~Kulaxizi and A.~Parnachev,
  ``Supersymmetry Constraints in Holographic Gravities,''
  arXiv:0912.4244 [hep-th].
}

\lref\DiFrancescoNK{
  P.~Di Francesco, P.~Mathieu and D.~Senechal,
  ``Conformal Field Theory,''
{\it  New York, USA: Springer (1997) 890 p}
}

\lref\MackJE{
  G.~Mack,
  ``All Unitary Ray Representations Of The Conformal Group SU(2,2) With
  Positive Energy,''
  Commun.\ Math.\ Phys.\  {\bf 55}, 1 (1977).
}

\lref\HeemskerkPN{
  I.~Heemskerk, J.~Penedones, J.~Polchinski and J.~Sully,
  ``Holography from Conformal Field Theory,''
  JHEP {\bf 0910}, 079 (2009)
  [arXiv:0907.0151 [hep-th]].
}

\lref\GrinsteinQK{
  B.~Grinstein, K.~A.~Intriligator and I.~Z.~Rothstein,
  ``Comments on Unparticles,''
  Phys.\ Lett.\  B {\bf 662}, 367 (2008)
  [arXiv:0801.1140 [hep-ph]].
}

\Title{\vbox{\baselineskip12pt
}}
{\vbox{\centerline{Energy Flux Positivity and Unitarity  in CFTs}
\vskip.06in
}}
\centerline{Manuela Kulaxizi${}^a$ and Andrei Parnachev${}^b$}
\bigskip
\centerline{{\it ${}^a$Department of Physics, University of Amsterdam }}
\centerline{{\it Valckenierstraat 65, 1018XE Amsterdam, The Netherlands }}
\centerline{{\it ${}^b$C.N.Yang Institute for Theoretical Physics, Stony Brook University}}
\centerline{{\it Stony Brook, NY 11794-3840, USA}}
\vskip.1in \vskip.1in \centerline{\bf Abstract}
\noindent
We show that in most conformal field theories the condition of the energy flux positivity,
proposed by Hofman and Maldacena, is equivalent to the absence of ghosts.
At finite temperature and large energy and momenta, the two-point functions
of the stress energy tensor develop lightlike poles.
The residues of the poles can be computed, as long as the only spin two conserved current,
which appears in the stress energy tensor OPE and
acquires nonvanishing expectation value at finite temperature, is the stress energy tensor.
The condition for the residues to stay positive and the theory to remain ghost
free is equivalent to the condition of positivity of energy flux.

\vfill

\Date{July 2010}


\newsec{Introduction}

\noindent Recently, considerable discussion was devoted
to the AdS/CFT correspondence for gravitational theories with
higher derivative interactions.
In particular, it has been observed that conformal field theories (CFTs)
dual to some of these models share an interesting property.
Namely, the requirement of causal propagation  of high energy modes at finite temperature \BriganteNU\
is equivalent to requiring the  positivity of energy flux \HofmanMaldacena.
(Positivity of energy flux and its equivalence to causality
has also been studied in
\refs{\BuchelTT\HofmanUG\deBoerPN\CamanhoVW\BuchelSK\deBoerGX\CamanhoHU-\MyersJV}.
See also \LatorreEA\ for earlier work in this direction and \ParnachevYT\
for a check of energy flux positivity in a number of interacting
superconformal theories.)
The causality constraints follow from the dispersion relation in the regime
where the frequency $w$ and momentum $q$ are much larger than the temperature $T$.
The positivity of energy constraints involve the two and three point functions of
the stress energy tensor, which can be determined from the singular terms in the OPE of stress energy tensor
with itself.
Hence, both sets of constraints follow from the high energy (UV) properties of the CFT,
and it is natural to ask whether there is a general argument which relates them.

In this paper we investigate this question from the field theoretic point of view\foot{
Earlier arguments for the positivity of energy flux in CFTs can be found in
an interesting paper \HofmanUG.}.
We start by considering a CFT where the only operator which takes an expectation
value at finite temperature is the stress energy tensor.
This condition seems natural for consistent CFTs defined by pure gravitational
theories where the only degree of freedom is a massless graviton in the bulk dual
to the boundary stress energy tensor.
We then compute the first non-trivial finite temperature correction to the two-point function of
the stress energy tensor in the regime of small temperatures, $q/T,w/T\gg 1$.
We show that whenever the flux positivity conditions of \HofmanMaldacena\ are
violated, the residue of the lightlike pole acquires a negative sign, signifying
the appearance of ghosts in the spectrum and violation of unitarity.

More precisely, the positivity of energy flux constraints in a four-dimensional CFT are
given by eq. (2.38) of \HofmanMaldacena:
\eqn\posenergy{\eqalign{
     &\CC_{tensor}=\left( 1-{t_2\over3}-{t_4\over15}\right)\ge 0 \cr
     &\CC_{vector}=\left( 1-{t_2\over3}-{t_4\over15}\right)+{t_2\over2}\ge 0 \cr
     &\CC_{scalar}=\left( 1-{t_2\over3}-{t_4\over15}\right)+{2t_2\over3}+{2t_4\over3}\ge 0 \cr
}}
where the subscript in $\CC$ corresponds to the properties
of the state with respect to rotation around the unit vector
which specifies the point on $S^2$ where the energy flux is measured \HofmanMaldacena.
The variables $t_2$ and $t_4$ can be expressed as
functions of the parameters $a,b,c$ that determine the two and three
point functions of the stress energy tensor.
They are given by eq. (C.12) in \HofmanMaldacena\foot{These expressions, as well as \posenergy,
have been generalized to six \deBoerPN\ and arbitrary \refs{\CamanhoVW,\BuchelSK} spacetime
dimensions.}
and upon substitution into \posenergy\ give rise to
\eqn
\posenergyc{\eqalign{
     & \CC_{tensor}=-{5 (7 a+2 b-c)\over (14 a-2 b- 5c)} \ge 0 \cr
     &\CC_{vector}= {10 (16 a+5 b-4 c)\over (14 a-2 b- 5c)}\ge 0 \cr
     &\CC_{scalar}= -{45 ( 4 a+2 b-c)  \over (14 a-2 b- 5c)} \ge 0 \cr
}}
In the next Section we compute the two-point function of stress energy
tensor in the regime $w/T,q/T\gg 1$.
There are three independent propagators which correspond to scalar, shear
and sound modes.
We show that the residues of the lightlike poles in scalar, shear, and
sound channels are equal (up to a positive numerical coefficient)
to $\CC_{tensor},\CC_{vector}$ and $\CC_{scalar}$ respectively.
This means that the energy flux positivity is related to the
absence of ghosts in the spectrum.
In Section 3 we discuss our results and outline some open questions.
We explain that the only case where our results can be modified involves
a conserved spin two current, other than the stress energy tensor,
which appears in the stress energy tensor OPE and takes an expectation value at finite temperature.
The only such case we are aware of involves a sum of decoupled CFTs.

\newsec{Flux positivity and ghosts}
\noindent
In this section we are going to compute the leading finite temperature
contribution to the short-distance behavior of the two-point function
of the stress energy tensor.
We start by considering the
OPE of the stress energy tensor $T_{\mu\nu}$ with itself, which takes the form \OsbornCR:
\eqn\tope{  T_{\mu\nu}(x) T_{\sigma\rho}(0)\sim {C_T \II_{\mu\nu,\sigma\rho}(x)\over x^{2 d}} +
            {\hat A}_{\mu\nu\sigma\rho\alpha\beta}(x) T_{\alpha\beta}(0)+\ldots     }
where $\II_{\mu\nu,\sigma\rho}(x)$ and  ${\hat A}_{\mu\nu\sigma\rho\alpha\beta}(x)$ are
known functions of $x$ which can be found in \OsbornCR.
In \tope\ we only keep the composites of the stress energy tensor in the right
hand side.
In principle, there can be other contributing terms, including relevant and marginal operators,
but we postpone
the discussion of those  until Section 3.
At this stage, we would just like to comment that consistent CFTs dual to pure gravitational theories,
whose operator content in this limit consists only of the stress energy tensor and its composites,
should be described by \tope. This is natural  from the point of view of gravity where the sole degree
of freedom is a massless graviton.
Interestingly,  studies of the superconformal algebra of the $\NN=4$
SYM in the strong coupling and large $N$ limit have also provided evidence in favor of
the OPE \tope\ \Anselmia.

To compute the two-point function, one takes the expectation
value of both sides in \tope.
Upon the Fourier transform to the momentum space, the first term in the right
hand side of \tope\ produces the usual Lorentz invariant (zero temperature)
two-point function,
\eqn\glinv{  G_0(k)\sim k^d \log k^2   }
where $k$ is the energy-momentum $d$-vector, and the suppressed index structure
is uniquely determined by Lorentz invariance and conformality.
We will be interested in the finite temperature correction to \glinv,
obtained by subtracting the zero temperature term
\eqn\tttdef{  \langle T_{\mu\nu}(x) T_{\sigma\rho}(0) \rangle_T =
                 \langle T_{\mu\nu}(x) T_{\sigma\rho}(0) \rangle - \langle T_{\mu\nu}(x) T_{\sigma\rho}(0) \rangle|_{T=0}  }
where the expectation value $\langle\ldots\rangle$ is taken at finite temperature $T$;
consequently $\langle\ldots\rangle_T$ denotes the finite temperature expectation value
with the  zero temperature contribution subtracted.
The leading contribution to $\langle T_{\mu\nu}(x) T_{\sigma\rho}(0) \rangle_T$
in the limit $k/T\gg1$ ($x T\ll 1$) comes  from the second term in \tope.
Contributions from less singular terms [denoted by the dots in \tope]
are suppressed by powers of $T/k$.
Since we are interested in the small temperature limit, it is sufficient to
compute \tttdef\ in Euclidean signature, use integral Fourier transform, and then
Wick rotate to Minkowski  space.
One can presumably recover interesting subleading finite temperature effects
by performing this calculation in the real time formalism.

In the following we will restrict our attention to the
$d=4$ case, although the generalization to arbitrary $d$ should
be straightforward.
We also choose the coordinates so that the spatial momentum
is oriented along the $x_3$ direction, $k=(w,0,0,q)$.
We start by considering the case  of transversely polarized $T_{\mu\nu}$:
\eqn\gdef{  G_{12,12}(w,q)_T= \int d^4 x  \langle T_{12}(x) T_{12}(0) \rangle_T e^{-i w x_0-i q x_3}   }
The leading contribution to $\langle T_{12}(x) T_{12}(0) \rangle_T$
can be written as
\eqn\ttlead{  \langle T_{12}(x) T_{12}(0) \rangle_T =
   \CC\, T^{4} \left( -3  {\hat A}_{121200}(x) +\sum_{i=1}^{3} {\hat A}_{1212ii}(x) \right)
                                                     +\ldots }
where $\CC$ is a positive real constant  and the dots denote terms suppressed by powers of $T/w,T/q$.
In deriving \ttlead\ we took the expectation value of \tope\ at finite temperature,
subtracted the zero-temperature term and used the nonzero expectation values of $T_{\mu\nu}$:
\eqn\exptcft{  \langle T_{00}\rangle = -3  \CC T^{4}, \qquad
                \langle T_{ii}\rangle =  \CC T^{4}, \;\; i=1,\ldots,3  }
where the familiar Lorentzian result $T_{\mu\nu}=\CC\,\times {\rm diag}[3 p,p,p,p]$, with $p$
denoting the pressure,
has been Wick rotated to Euclidean signature.
Hence, to compute $G_{12,12}(w,q)_T$ we only need to know ${\hat A}_{1212\alpha\beta}(x)$.
Fortunately, it has already been computed: it is given by
eq. (6.38) in \OsbornCR.
There are two terms in ${\hat A}_{1212\alpha\beta}(x)$ computed in \OsbornCR:
the power-like term $\sim x^{-4}$ and  a delta function term $\sim \delta^{(4)}(x)$.
We will be interested in the former, since the latter produces a constant upon
the Fourier transform \gdef.

The first step in performing the Fourier transform \gdef\ on
the power-like term in \ttlead\
involves integrating over the transverse coordinates $x_1,x_2$.
The only contributions that survive this integration come from
${\hat A}_{121200}(x)$ and ${\hat A}_{121233}(x)$.
The expression for $ G_T(w,q)$ takes a  simple form:
\eqn\gb{  G_{12,12}(w,q)_T= \CC_{tensor}\, \int d x_0 d x_3  e^{-i w x_0-i q x_3}  {x_3^2-x_0^2\over (x_0^2+x_3^2)^2} }
where we neglected an overall positive numerical factor, and
$\CC_{tensor}$ is defined in the first line of \posenergyc.
Remarkably, the correlator is proportional to the combination of
parameters required to be positive by the condition of energy flux positivity.

To complete the integration in \gb\ we make use of the following trick.
We exponentiate the denominator in \gb\ via
\eqn\expden{  {1\over (x_0^2+x_3^2)^2} = \int_0^\infty ds s e^{-s (x_0^2+x_3^2)}   }
We then substitute \expden\ into \gb\ and perform the Gaussian integration over
$x_0,x_3$.
The result (up to a positive and real overall constant) is
\eqn\gc{   G_{12,12}(w,q)_T= \CC_{tensor} \int_0^{\infty} ds {q^2-w^2\over s^2} e^{-{q^2+w^2\over s }}  }
We can now perform the integration over $s$ and Wick rotate to
Lorentzian signature to obtain
\eqn\gd{   G_{12,12}(w,q)_T= \CC_{tensor} {w^2+q^2\over w^2-q^2}      }
There is a lightlike pole whose residue changes when $\CC_{tensor}$ changes sign.
We see that the first line in \posenergy\ is equivalent to the absence of ghosts
in the scalar channel.

The situation is similar for the other polarizations of the stress energy tensor.
Recall that generally there are three independent components in the two-point
function of the stress energy tensor: scalar, shear and sound.
(Such a separation is a result of imposing Ward identities on
all consistent tensor structures which can appear in the two-point functions.)
The corresponding correlators take the form (see e.g. \KovtunEV):
\eqn\modes{\eqalign{
          G_{12,12}(k)&={1\over2} G_3(w,q)\cr
          G_{10,13}(k)&=-{1\over2}{w q\over w^2-q^2} G_1(w,q)\cr
          G_{00,00}(k)&={2 q^4\over 3(w^2-q^2)^2} G_2(w,q)\cr
}}
The correlator computed in \gd\ corresponds to the scalar mode, $G_{12,12}(k)$.
The computation in the shear channel involves
\eqn\ttleadb{  \langle T_{10}(x) T_{13}(0) \rangle_T =
   \CC\, T^4 \left( -3 {\hat A}_{101300}(x) +\sum_{i=1}^{3} {\hat A}_{1013ii}(x) \right)
                                                   }
Now all ${\hat A}_{1013ii}(x)$ contribute, and after integration over $x_1,x_2$,
we have
\eqn\gsheart{ G_{10,13}(k)_T=  \int d x_0 d x_3 e^{-i w x_0-i q x_3}
      { (64 a{+}14 b{-}19 c) x_0^3 x_3-(64 a{+}26 b{-}13 c) x_0 x_3^3\over(14 a-2 b-5 c)(x_0^2+x_3^2)^2}   }
where we omitted an overall positive numerical factor.
We can use the exponentiation of the denominator [similar to  \expden] once again to obtain
\eqn\shearid{ \int d x_0 d x_3  e^{-i w x_0-i q x_3} {x_0^3 x_3\over (x_0^2+x_3^2)^3} ={wq (13 w^2-3 q^2)\over (q^2+w^2)^3  }}
Substituting \shearid, and an expression obtained from \shearid\ by the simultaneous
interchange $(x_0\leftrightarrow x_3, w\leftrightarrow q)$,   into \gsheart\ and \modes\ and performing a  Wick rotation, we obtain
\eqn\shearres{   G_1(w,q)= - {(512 a+130 b-143 c) w^2+ (512 a+190 b- 113 c) q^2
       \over (14 a-2 b-5 c) (w^2-q^2)}    }
The residue of the pole at $w^2=q^2$ in the limit $q,w\ra\infty$ is proportional to
$\CC_{vector}$, up to a positive number.
The condition for the absence of ghosts in the shear channel is therefore equivalent
to the second line in \posenergy.
Finally, we expect the last line in \posenergy\ to correspond
to the absence of ghosts in the sound channel.
This can be verified through the computation of $ G_{00,00}(k)$.
The part that contains the pole takes the form
\eqn\gsound{\eqalign{ G_{00,00}(k)_T&=   \int d x_0 d x_3 e^{-i w x_0-i q x_3}\times \cr
     &  { ({-}14 a {+} 11 b {+} 5 c) x_0^6 + (20 c{-}37 (2 a + b)) x_0^4 x_3^2 {+}
 3 (18 a {-} b {-} 5 c) x_0^2 x_3^4 + 3 (6 a {-} b {-} 2 c) x_3^6\over(14 a-2 b-5 c)(x_0^2+x_3^2)^4}   \cr}
}
This integral can be computed using the same techniques as
above.
The result, up to a positive real numerical factor, is
\eqn\soundres{
G_2(w,q){=}
{({-}68 a {+} 32 b {+} 23 c) q^6 {+}
 3 (52 a {+} 16 b {-} 15 c) q^4 w^2 {+} (116 a {-} 32 b {-}
    35 c) q^2 w^4 {-} (76 a {-} 16 b {-} 25 c) w^6
 \over -(14 a-2 b-5 c) (w^2-q^2)}       }
which implies that the residue in $G_2(w,q)$ at $w^2=q^2$ is
proportional to $C_{scalar}$ in the limit $w,q\ra\infty$ and the absence of ghosts is equivalent
to the third line in \posenergy.

\newsec{Discussion}

\noindent We showed that the positivity of energy flux in a CFT, described by \posenergy,
is equivalent to the absence of ghosts\foot{The positivity of the residues of the poles 
can be shown to be equivalent to the absence of ghosts in a number of ways.  
For example, the corresponding Wightman functions must be positive.
This statement translates into reflection positivity of the Euclidean theory.}.
Their positivity at finite temperature\foot{It would be interesting
to understand how our work is related to the recent discussion of unitarity
in the context of AdS/CFT \HeemskerkPN.},
provided the only operator with non-vanishing expectation value which
appears in the right hand side of \tope\ is the stress energy tensor itself.
An irrelevant operator with nonvanishing expectation value would not alter the
discussion, since its contribution would be further suppressed by positive powers
of $T/q,T/w$.
However a marginal operator with nonvanishing finite temperature expectation
value would contribute to $G(w,q)_T$ at the same order, while the corresponding
contribution from a relevant operator would dominate in the low temperature
limit.
We only need to consider primary operators, since the descendants  involve
$\p_\mu({\rm something})$, and their expectation
values vanish in the CFT at finite temperature.
First, consider a scalar operator $\Phi$, other than the identity, with conformal
dimension $\Delta$ between one and four and a nonvanishing expectation
value.
The Lorentz invariance of the OPE, together with the fact that\foot{
This scaling follows from $T$ being the only dimensionful parameter in the theory.} 
$\langle \Phi\rangle\sim T^{\Delta}$ does not break Lorentz invariance,
implies that it does not contribute to the poles in $G_i(w,q)$.
This is because the contribution from such a scalar to $G_i(w,q)$ would be proportional to
$T^{\Delta} (k^2)^{2-{\Delta\over2}}$, which is non-singular for $\Delta\leq4$.
We have also explicitly checked this using the $TT\sim \Phi$ OPE.
As evident from \glinv\ there can be corrections to the scaling which go as
$\log k^2$, but these do not affect the pole structure.

The only relevant operator whose expectation value could break Lorentz invariance
is a vector.
Rotational invariance implies that only the time component of a vector could
take an expectation value.
One can then use a rotation by  $\theta=\pi$ in the $x^0-x^1$
plane, which is still a symmetry of the finite temperature theory, to
deduce that the time component of a vector has to vanish as well,
as long as this residual symmetry is not spontaneously broken.
Hence, the only operator which can spoil the correspondence  between
the positivity of energy flux and the absence of ghosts is a traceless
symmetric spin-2 conserved current, which is not proportional to
$T_{\mu\nu}$.
(The trace part does not violate Lorentz invariance, while a non-conserved
spin-2 operator is necessarily irrelevant \MackJE; see also e.g. \GrinsteinQK\ for
a recent discussion.)\foot{The arguments
in this paragraph are due to  J. Maldacena.}

Such an operator, which we denote by $X_{\mu\nu}$ below, generates a copy of the conformal algebra, and can in principle
appear in the $T_{\mu\nu}(x)T_{\alpha\beta}(0)$ OPE.
The simplest (and the only known to us) example is the case of two\foot{The case
of more than two decoupled CFTs is a simple generalization of this.}
decoupled CFTs, whose stress energy tensors
are denoted by $T^{(1)}_{\mu\nu}$ and $T^{(2)}_{\mu\nu}$.
In this case the structure of the OPE implies that, in addition to
$T_{\mu\nu}=T^{(1)}_{\mu\nu}+T^{(2)}_{\mu\nu}$,
there is another linear combination of $T^{(1)}_{\mu\nu}$ and $T^{(2)}_{\mu\nu}$
which appears in the right hand side of the OPE \tope.
Of course, in this case it is possible to diagonalize the OPE,
so that $T^{(1)}_{\mu\nu} T^{(1)}_{\mu\nu}\sim T^{(1)}_{\mu\nu}$ and
$T^{(2)}_{\mu\nu} T^{(2)}_{\mu\nu}\sim T^{(2)}_{\mu\nu}$.
It is an interesting question whether this is always possible
in general.
A curious case of a spin-2 operator which is not proportional
to $T_{\mu\nu}$ occurs in two spacetime dimensions for the $SU(2)_8$
WZNW theory\foot{This example is due to N. Seiberg.}.
This model admits a $D_8$ modular invariant which contains a $\chi_8 \chi_0^*$
term (in the notations of \DiFrancescoNK).
Hence, the spectrum contains a primary operator of dimension (2,0), other than the stress energy tensor.
This operator, however, does not appear in the stress energy tensor OPE.

Given the appearance of $X_{\mu\nu}$ in the OPE \tope, one may
ask whether the stress energy tensor Ward identities may be helpful in
establishing that $\langle X_{\mu\nu}\rangle=0$.
In particular, the conformal Ward identity implies
\eqn\cwid{  \langle T^\mu_{\mu}(x) T_{\alpha\beta}(0)\rangle =  4 \langle T_{\alpha\beta}(0)\rangle \, \delta^{(4)}(x) }
This may naively seem to imply that $\langle X_{\mu\nu}\rangle=0$.
This however is not the case, since only the $\delta^{(4)}(x)$ term in the
$T_{\mu\nu}(x)T_{\alpha\beta}(0)$ OPE produces the nonvanishing result upon contraction with
$\eta^{\mu\nu}$ and taking expectation value at finite temperature.
We neglected these terms in our computation in Section 2,
since they do not contribute to the residues.
The stress energy tensor Ward identities guarantee that the coefficients of
such terms ensure \cwid, while analogous terms for the $X_{\mu\nu}$
in the $T_{\mu\nu}(x)T_{\alpha\beta}(0)$ OPE  are absent.
Hence, a possible contribution from the $\langle X_{\mu\nu}\rangle$
to the two-point function of the stress energy tensor may exist,
although it vanishes upon contraction with $\eta^{\mu\nu}$.

One may wonder whether there exists some rotationally and translationally invariant state,
where the only spin-2  operator with
nonvanishing expectation value  is $T_{\mu\nu}$,
and the value of the pressure, given by $\langle T_{ii}\rangle$, is positive.
In this case the results of the previous section would go through, because
these were the only assumptions involved.
(Of course, a finite temperature state described by the density
matrix $\rho\sim e^{-H/T}$ satisfies these assumptions.)
It would be interesting to see if one can find such state, perhaps
subject to the constraints discussed in \LatorreEA.

In CFTs dual to pure gravitational theories, the composites of the stress
energy tensor should be the only operators appearing in \tope.
It is therefore not surprising that the correspondence between
the energy flux positivity and the violation of causality
has been first observed in CFTs dual to Gauss-Bonnet
\refs{\BriganteNU\HofmanMaldacena\BuchelTT\HofmanUG\deBoerPN\CamanhoVW-\BuchelSK}
and Lovelock theories of gravity \refs{\deBoerGX,\CamanhoHU} with negative
cosmological constant.
In these models, once the parameters of the theory are taken
outside of the flux positivity region, a set of tachyonic
quasinormal modes appears at finite temperature.
The corresponding  states in the dual CFT are stable in the limit $w/T,q/T\ra\infty$,
and have velocities which vary from unity to a finite number $c_g>1$.
These states form a continuum in the $w/T,q/T\ra\infty$ limit.
Presumably, our CFT calculation observes the lower edge of this
continuum of states and predicts that these states are also ghosts,
in addition to being tachyonic.
One may wonder  whether this picture is compatible with the results
from gravity in the regime where the inequalities in \posenergy\ are satisfied.
In this case there is no  metastable state from the gravity point of view.
However, as the results of \KovtunEV\ suggest, the imaginary part of the
pole goes to a constant in the limit $q/T,w/T\gg 1$ and therefore is
not visible at leading order, which is indeed  consistent with our results.

It would be interesting to verify this picture directly.
It would also be interesting to understand better the situation
with the quasi-topological gravity that has been recently discussed in
\refs{\OlivaEB,\MyersRU,\MyersJV}.
Exploring consequences of our results in the theories at finite
size is an interesting direction of research.
The generalization of the results of this paper to
the supersymmetric case should be straightforward.
One would need to consider the OPE of the R-current operator with itself.
The results should be consistent with vanishing $t_4$ in the supersymmetric case \KulaxiziPZ.
Finally, we would like to point out that the discussion
of this paper should also apply to field theories which are
defined via perturbing CFTs by relevant operators and subsequent
RG flow.
This is because the discussion concerns the UV limit, where
one necessarily recovers the  original CFT.

\bigskip
\bigskip

\noindent {\bf Acknowledgements:}
We thank A. Buchel, F. Dolan, J. Harvey, D. Kutasov, J. McGreevy, J. Minahan, M. Rangamani, L. Rastelli, K. Skenderis, E. Verlinde, 
L. Yaffe and especially K. Papadodimas for useful discussions and correspondence.
We are especially grateful to J. Maldacena for numerous  explanations, suggestions and
comments on the manuscript.
We thank Aspen Center for Physics and NORDITA, where parts of this work
have been completed for hospitality.

\footatend\vfill\supereject\immediate\closeout\rfile\writestoppt
\baselineskip=14pt\centerline{{\bf References}}\bigskip{\frenchspacing%
\parindent=20pt\escapechar=` \input refs.tmp\vfill\eject}\nonfrenchspacing
\end